\documentclass[prb,twocolumn,preprintnumbers,amsmath,aps]{revtex4}
\usepackage{graphicx,hyperref}% Include figure files
\usepackage{bm}% bold math
\usepackage{subfigure}% bold math
\usepackage{xcolor}

\begin{document}

\title{A perspective on the fragility of glass-forming liquids}

\author{Christiane Alba-Simionesco} \email{christiane.alba-simionesco@cea.fr}
\affiliation{University Paris-Saclay, CEA, CNRS, Laboratoire L\'eon Brillouin, 91191, Gif-sur-Yvette, France}

\author{Gilles Tarjus} \email{tarjus@lptmc.jussieu.fr}
\affiliation{Laboratoire de Physique Th\'eorique de la Mati\`ere Condens\'ee, CNRS-UMR7600, Sorbonne Universit\'e, 4 Place Jussieu, 75005 Paris, France}

\begin{abstract}
We discuss possible extraneous effects entering in the conventional measures of ``fragility'' at atmospheric pressure that may obscure a characterization of the 
genuine super-Arrhenius slowdown of relaxation. We first consider the role of density, which increases with decreasing temperature at constant pressure, and then the 
potential influence of the high-temperature dynamical behavior and of the associated activation energy scale. These two effects involve both thermodynamic parameters 
and the strength of the ``bare'' activation energy reflecting the specific bonding between neighboring molecules. They vary from system to system with, most likely, 
little connection with any putative collective behavior associated with glass formation. We show how to scale these effects out by refining the definition of fragility and 
modifying the celebrated Angell plot. We dedicate this note to our great and so inspiring friend, Austen Angell, and associate in this tribute another dear colleague who 
died too soon, Daniel Kivelson.
\end{abstract}
\keywords{glass transition 1 $|$ fragility 2 $|$ isochoric fragility 3 $|$ Arrhénius range 4 $|$ lengthscale  5$} 

\maketitle

\section{Introduction}

\begin{figure}[t]
\includegraphics[width=0.99\columnwidth]{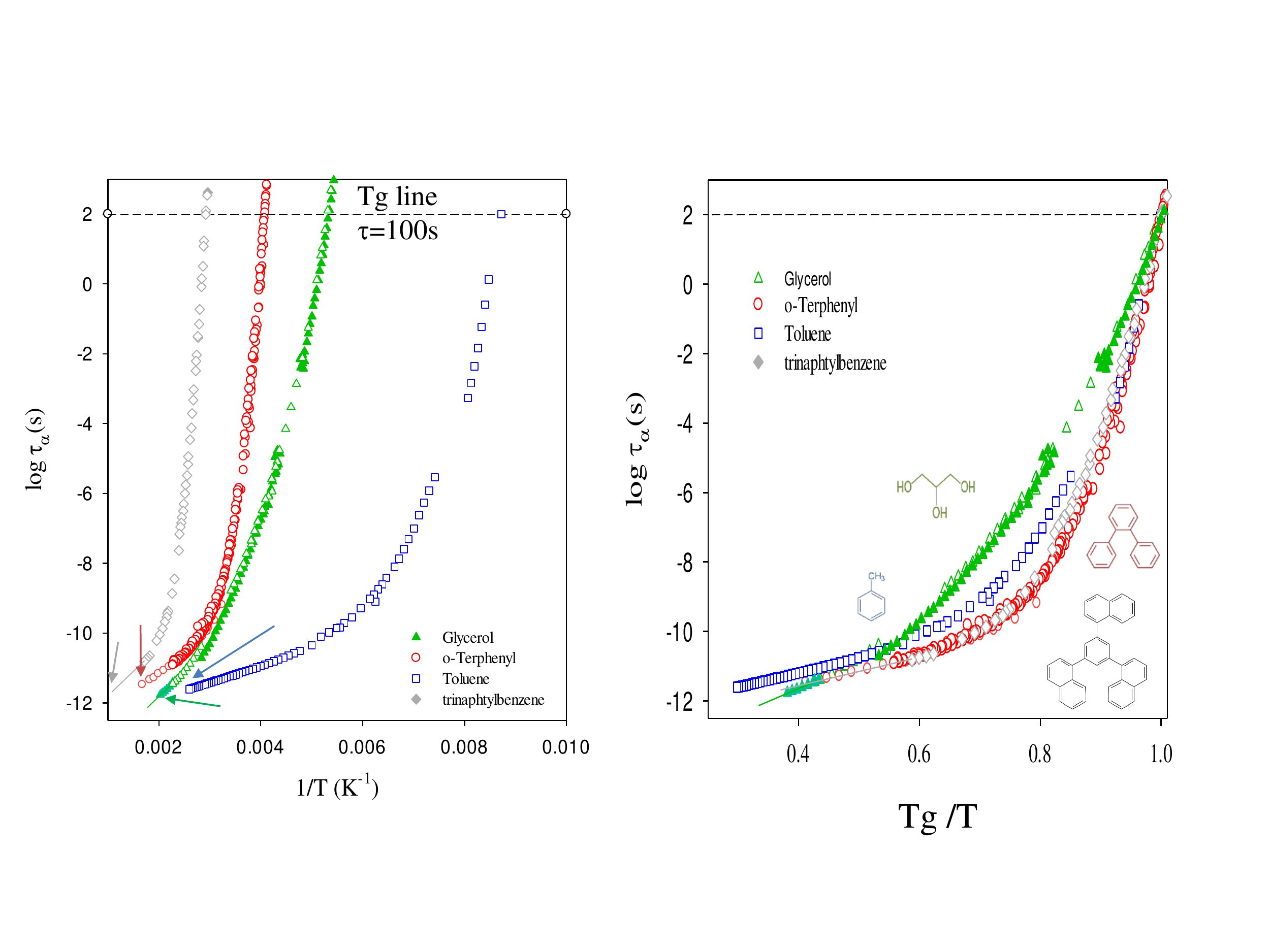}
\caption{Temperature dependence of the $\alpha$-relaxation time and the viscosity of several molecular glass-forming liquids at atmospheric pressure. 
Left: $\log_{10}(\tau(P_{\rm atm},T)/{\rm s})$ versus $1/T$ in the whole liquid range from the boiling point (indicated by an arrow) to the glass transition $T_g$ defined for 
$\tau=100s$ (color code explained in the figure). 
Right: The celebrated Angell plot in which the temperature is scaled by $T_g$. The molecular structure of the liquids is also shown. One can observe that the fragility measured 
as the slope of $\log_{10}(\tau(P_{\rm atm},T)/{\rm s})$  plotted versus $T_g/T$ evaluated at $T_g$ is quite similar for the three aromatic liquids, with a steepness index 
$m_p\approx77-82$, while it is smaller for glycerol, with $m_p\approx40-50$. (Data mostly from [\onlinecite{rossler}]; additional data for glycerol from [\onlinecite{GlyPatm}],  
for  $\alpha \alpha \beta$-tris-naphthylbenzene from [\onlinecite{TNB}], for toluene from [\onlinecite{tol}], and for o-terphenyl from [\onlinecite{oTP}].)}	
\label{Fig_Tg-scaling}
\end{figure}

Summer 84. In the basement of the Chemistry Building at Purdue University, Austen's lab, there was a young and dynamic team of students and post-docs, 
coming  from all around the world to have a chance to work with Austen on the glass phenomenology. We were all busy answering Austen's questions and providing the 
data he required by all the means of communication available at the time, telephone (regardless of time difference with Rome, Paris, or Washington) or fax.  Viscosity data 
from the library and the Beilstein database, new values of $T_g$ or of the heat capacity... he needed all of this for his talk at the Blacksburg Workshop on Relaxations in Complex 
Systems where he first introduced the strong-fragile classification [\onlinecite{angell85,angell85_bis}]. (At the same time Austen was also trying to extend the range of fragilities by vitrifying 
simple and supposedly non-glass-forming liquids through soft confinement in microemulsions [\onlinecite{dubochet}].) Austen proposed the concept of 
fragility [\onlinecite{angell85,angell85_bis,angell91,angell_steepness}] to characterize how quickly transport coefficients and relaxation times increase as one cools a 
glass-forming system down 
to the glass transition temperature: this is illustrated in Fig. \ref{Fig_Tg-scaling} for archetypical glass-forming liquids. We, students and post-docs, did not realize at the 
time the impact of the concept, but it strongly influenced and even determined our future research, at least for me (CAS).

Fragility focuses on a generic, striking, property of the dynamical slowdown in most glass-forming molecular liquids, polymers and inorganic materials, namely the fact that the temperature dependence of the dynamical properties is stronger than an Arrhenius one. This approach was further applied to other classes of systems such as metallic liquids 
or even spin glasses [\onlinecite{Souletie}]. Later, guided by the Adam-Gibbs approach relating the increase of the relaxation time to the decrease of the configurational entropy [\onlinecite{adam-gibbs65}], Austen formulated the notion of thermodynamic fragility [\onlinecite{angell_thermo,martinez-angell}], which quantifies his former intuition concerning the correlation 
with the loss of degrees of freedom and the jump of the heat capacity $C_{p}$  at $T_g$ [\onlinecite{angell85,angell85_bis}].
\\

Fall 1997. In UCLA we (GT, CAS) were working with Daniel Kivelson, another great physical chemist and close friend gone too soon, on the glass transition problem 
and on the role of density and temperature in the slowing down of relaxation [\onlinecite{ferrer98,ferrer98_bis}]. Daniel's intuition was that Austen's scaled Arrhenius plot 
obtained from data 
at constant (atmospheric) pressure should be complemented by information allowing one to disentangle the respective contributions of density and temperature. This first led 
him to envisage an extended version of Austen's classification, based on constant-density plots and distinguishing a fragile-versus-nonfragile axis that describes 
the degree of super-Arrhenius behavior and a strong-versus-weak axis that accounts for the value of the  high-temperature Arrhenius effective activation 
energy [\onlinecite{ferrer_extension}]. We will build up on these ideas below. 
\\

Why is the concept of fragility useful beyond its value as a classification scheme? In trying to rationalize and sort out the phenomenological observations on 
glass-forming systems, a number of correlations have been empirically investigated between measures of fragility and other material-specific quantities. Such 
quantities may be (i) thermodynamic, such as the amplitude of the 
heat-capacity jump at the glass transition [\onlinecite{angell85,angell85_bis,angell91}] or the steepness of the decrease of the configurational 
entropy with temperature [\onlinecite{angell_thermo,martinez-angell,richert-angell98,sastry01,stevenson-wolynes05,ludo_AG}], (ii) associated with short-time or glass properties, 
such as the relative intensity of the boson peak [\onlinecite{sokolov_boson}], the mean square displacement at $T_g$ [\onlinecite{ngai_MSD}], the ratio of elastic 
to inelastic signal in the X-ray Brillouin spectra [\onlinecite{scopigno_glass}], the Poisson ratio [\onlinecite{novikov_poisson}], or the temperature dependence of 
the elastic shear modulus $G_\infty$  in the viscous liquid [\onlinecite{dyre_elastic-review}], or (iii) characteristic of the relaxation slowdown in the supercooled liquid 
regime, such as the ``stretching'' of the time-dependent $\alpha$-relaxation functions [\onlinecite{bohmer_stretching}] or the extent of the 
dynamical heterogeneities [\onlinecite{capaciolli_dynhet}].

These empirical correlations may serve as motivations for theoretical approaches [\onlinecite{lubchenko-wolynes}]. However, beyond the obvious caveat that correlation does not 
imply causation, extracting the various quantities from experimental data (including measures of the fragility) usually involves large error bars and many of the proposed 
correlations have been challenged: see, e.g., [\onlinecite{mckenna00,berthier-ediger}]. Furthermore, fragility involves a variation with temperature that a priori 
depends on the thermodynamic path chosen, such as constant-pressure (isobaric) versus constant-density (isochoric) conditions. On the other hand, many quantities 
that have been correlated to fragility only depend on the thermodynamic state at which they are considered (e.g., the stretching parameter and many quantities 
measured at $T_g$). It therefore appears desirable to look for a characterization of the liquid fragility that reflects as much as possible the intrinsic 
super-Arrhenius temperature dependence of the relaxation time and the potential growing collective character that one often associate with 
it [\onlinecite{BB_review,tarjus_review,lubchenko-wolynes}]. In this paper we review and discuss steps in this direction. We focus on molecular liquids, with additional 
considerations on glass-forming polymers and ionic liquids when possible, but we leave out the whole class of network-forming systems that are generically on the 
strong side of Angell's classification (for a recent account, see [\onlinecite{sidebottom19}]). We also leave out the phenomenon of the fragile-to-strong transition or crossover 
first proposed by Austen Angell and coworkers to rationalize the behavior of liquid water [\onlinecite{angell_thermo}] and later observed in some multi-component 
metallic glass-formers and network-forming glasses [\onlinecite{zhang_10,lucas_21}].
\\

\section{Density scaling of dynamics and isochoric fragility}
\label{sec_density}

In the course of our work with Daniel Kivelson concerning the respective influence of temperature and density on the slowdown of dynamics in supercooled 
liquids, we discovered that the effect of density on the $\alpha$-relaxation time $\tau(\rho, T)$ and on the viscosity $\eta(\rho,T)$ in a variety of 
glass-forming liquids could be scaled out by considering the single combination of a density-dependent activation energy scale $e(\rho)$ divided by the 
temperature [\onlinecite{alba02}],
\begin{equation}
\label{eq_rho_scaling}
\log_{10}(\frac{\tau(\rho, T)}{\tau_\infty})= \mathcal F(\frac{e(\rho)}T)
\end{equation}
where $\tau_\infty$ is a (possibly material- and probe- dependent) constant and $\mathcal F(X)$ is a material-dependent but state-point 
independent function, or alternatively, when written in terms of an effective activation energy, 
\begin{equation}
E(\rho,T)= e(\rho) \mathcal G(\frac{e(\rho)}T)
\end{equation}
with $\ln(\tau(\rho, T)/\tau_\infty)=E(\rho,T)/T$ and the functions $\mathcal F$ and $\mathcal G$ related through $\mathcal F(X)=X \mathcal G(X)/\ln(10)$. 
In the above equations the Boltzmann constant $k_B$ has been set equal to 1.

The property is exact for models of monodisperse soft spheres interacting through a power-law pair potential, $v(r)\propto 1/r^n$ [\onlinecite{hansen-macdo}], for which 
$e(\rho)\propto \rho^{n/3}$. It was first noticed by T\"olle et al.[\onlinecite{tolle98}] in their description of inelastic incoherent neutron scattering data on o-terphenyl 
under pressure through a variable $\rho^{4}/T$.

We further investigated the consequences of this finding in Ref. [\onlinecite{tarjus04,alba_lille}] and also applied it to glass-forming 
polymers [\onlinecite{aude_thesis,alegria04}]. The density scaling was confirmed by other groups [\onlinecite{dreyfus03,dreyfus04,casalini-roland04}] and, since then, has been 
checked for a broad span of glass-forming liquids and polymers [\onlinecite{roland-paluch_review}].

\begin{figure}[t]
\includegraphics[width=0.99\columnwidth]{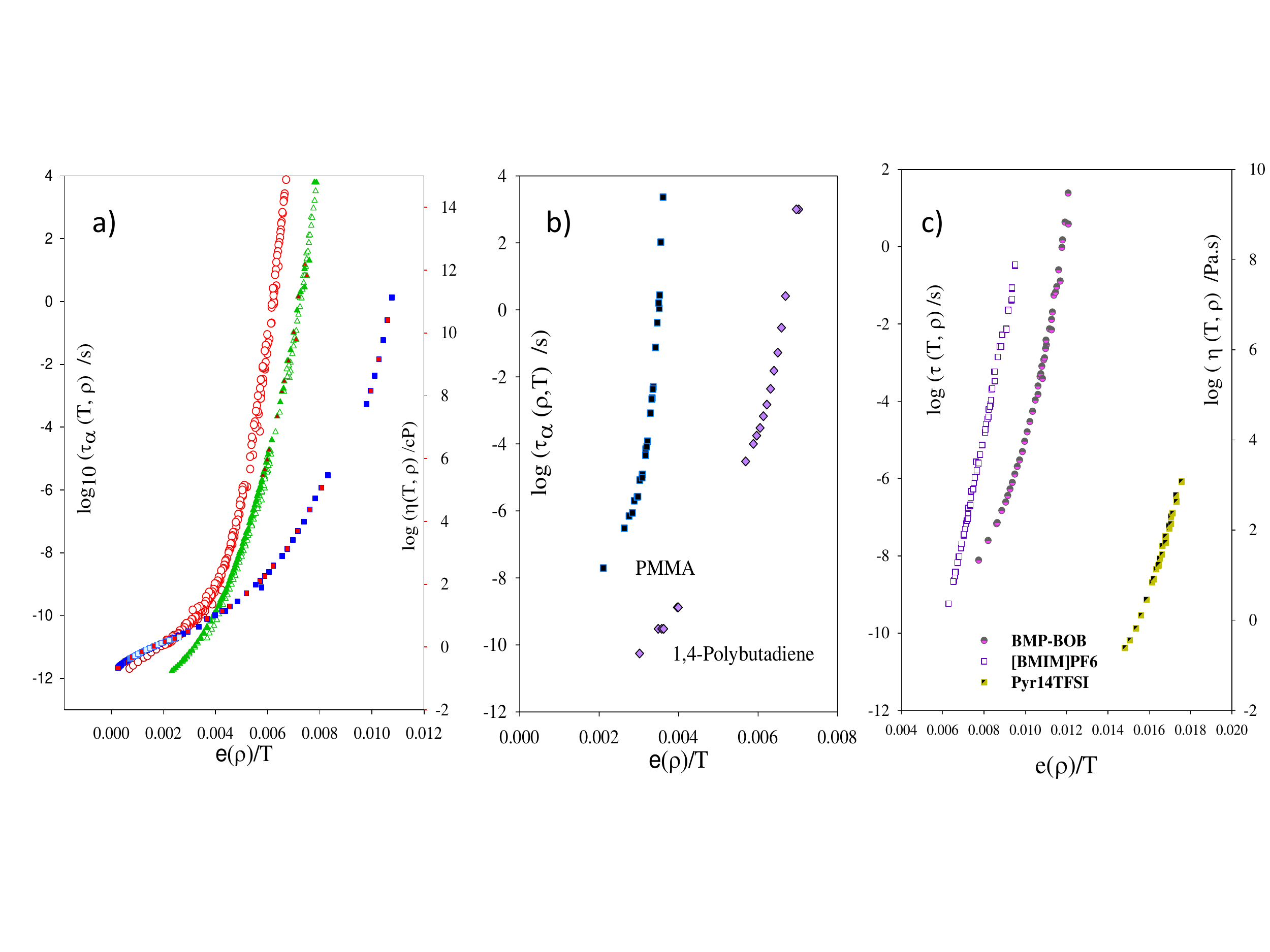}
\caption{Density scaling of the $\alpha$-relaxation time or the viscosity for several glass-forming molecular liquids studied at atmospheric pressure and high pressure: $\log_{10}(\tau(\rho,T)/{\rm s})$ versus the scaling variable $e(\rho)/T$, as explained in the text. (a) molecular liquids: o-terphenyl [\onlinecite{dreyfus03,tarjus04}], 
glycerol [\onlinecite{GlyHP}] and toluene 
(same color code as Fig. \ref{Fig_Tg-scaling}); (b) polymers: 1, 4-polybutadiene and PMMA [\onlinecite{aude_thesis,alegria04,PMMA}]; (c) ionic liquids: 
[BMIM]PF6 [\onlinecite{[BMIM]PF6}], BMP-BOB [\onlinecite{(BMP-BOB)}], and Pyr14TFSI [\onlinecite{Pyr14TFSI}]. Note that the range of variation is of course much 
larger for molecular liquids. If one describes the energy scale $e(\rho)$ by a power law, $e(\rho)\sim \rho^x$, the exponent is for instance  found to be $x\approx4$ for 
o-terphenyl, $2.7$ for 1,4-polybutadiene, and $1.3$ for PMMA;  but for glycerol it varies from $1.3$ to $1.8$ depending on the pressure and  relaxation-time ranges considered 
(in [\onlinecite{casalini_xrho}] it was even found to vary from $0.8$ to $3.5$); for ionic liquids,  $x\approx 2.9$  ([BMIM]PF6), $3.7$ (BMP-BOB), and $2.8$ (Pyr14TFSI). This 
is discussed in the text.}
\label{Fig_density-scaling}
\end{figure}

The density scaling of the relaxation time and the viscosity is illustrated from data collected for all available $(P,T)$ state points for several glass-forming 
molecular liquids (from the boiling point to $T_g$), ionic liquids, and polymers in Fig. \ref{Fig_density-scaling}.
\\

The physical interpretation of this density scaling is still debated. A simple parametrization of the energy scale $e(\rho)$ is via a power-law form, $e(\rho)\sim \rho^x$ with 
$x$ an exponent $>0$ [\onlinecite{tolle98,aude_thesis,alegria04,dreyfus03,casalini-roland04}]. Guided by the result for soft spheres interacting through a repulsive power-law 
potential, one may tentatively relate this behavior to the short-range repulsive component of the intermolecular potential [\onlinecite{tolle98,dreyfus03,casalini-roland04}]. However, 
this interpretation fails in many systems where an intrinsic energy scale (torsion energy, hydrogen bonding, etc.) comes into play and prevent the mere use of a scale-free 
density dependence[\onlinecite{alba_lille}]. It was also shown that even in a specific class of simple liquids, dubbed Roskilde-simple by J. Dyre and 
coworkers [\onlinecite{dyre_roskilde}], the 
functional dependence of $e(\rho)$ is more complicated and can, for instance, take the form of a combination of two different power laws in the case of the Lennard-Jones 
liquid [\onlinecite{dyre_LJ}]. Furthermore, we found that for some liquids for which a large enough domain of pressure and density is experimentally available, forcing a 
description of $e(\rho)$ in terms of a power law requires to make the exponent $x$ density dependent [\onlinecite{dalle-ferrier_sendai,niss07}], which undermines the 
fundamental nature of the power-law scaling. (This was later confirmed by  another group [\onlinecite{casalini_xrho}].)

If there is no clear thermodynamic interpretation of the density scaling, it can nonetheless be related to a dynamical property of glass-forming liquids at high temperature. 
When data is available at high enough temperature, an effective Arrhenius behavior is generally found to rather well describe the temperature dependence of the relaxation 
time and the viscosity [\onlinecite{alba02,tarjus04}], 
\begin{equation}
\label{eq_Arrhenius_rho}
\ln(\frac{\tau(\rho,T)}{\tau_\infty}) \approx \frac{E_\infty(\rho)}T,
\end{equation}
and it is then easy to see that $e(\rho)$ is then proportional to the Arrhenius effective activation energy $E_\infty(\rho)$. Such an identification is unfortunately not possible 
for glass-forming polymers and for other systems such as ionic liquids for which the range of accessible thermodynamic states is too limited and does not allow a proper 
determination of a high-temperature Arrhenius regime.

Above all it is important to stress that the density scaling of the dynamics is an approximate one (except for power-law interaction potentials in monodisperse systems) and 
is moreover specific to supercooled liquids and polymers. For instance, it does not apply to soft-condensed matter and granular systems with very short-ranged truncated 
potentials that undergo a jamming transition: see [\onlinecite{berthier-tarjus,berthier-tarjus_bis}].
\\

The existence of the (empirical) density scaling of the dynamics allows one to address the problem raised by Daniel Kivelson, that fragility measured at constant pressure 
includes not only the intrinsic effect of temperature but also the influence of the increasing density. To get around this, one should use a constant-density (``isochoric'') fragility in 
place of the standard ``isobaric'' one. Experimental data however are not collected under isochoric conditions and this makes the general use of the isochoric fragility more difficult. 
A major simplification comes from the density scaling. It is indeed easy to show that, whatever its precise definition, the 
isochoric fragility is independent of density as a consequence of the scaling in $e(\rho)/T$ [\onlinecite{tarjus04,alba_lille}].

Consider the generic isochoric steepness index defined for a fixed value $\tau$ of the $\alpha$-relaxation time,
\begin{equation}
m_\rho(\rho,\tau)=\frac{\partial \log_{10}(\tau(\rho,T)/\tau_{\infty})}{\partial(T_\tau(\rho)/T)}\Big \vert_{T=T_\tau(\rho)}
\end{equation}
where $T_\tau(\rho)$ is the temperature at which $\tau(\rho,T)=\tau$. Then, by using Eq. (\ref{eq_rho_scaling}), one has
\begin{equation}
m_\rho(\rho,\tau)=X_\tau \mathcal F'(X_\tau)
\end{equation}
where $X_\tau=e(\rho)/T_\tau$ and a prime denotes a derivative with respect to the argument 
of the function. Obviously, since by construction $\mathcal F(X_\tau)=\log_{10}(\tau/\tau_{\infty})$ and $\mathcal F(X)$ is state-point independent, 
$X_\tau$ depends on the value of the 
reference time $\tau$ but is independent of density. As a result, $m_\rho(\rho,\tau)\equiv m_\rho(\tau)$ is independent of density. A 
specific choice of $\tau$ of the order of $100$ s (or of viscosity $\eta\sim10^{11-13}$ Poise) corresponds to the usual definition of 
the glass transition temperature $T_g(\rho)$. This shows as a particular case that the isochoric steepness index of a glass-former is constant 
along the glass transition line, but this is true along any ``isochronic'' (\textit{i.e.}, constant relaxation time or constant viscosity) line.
\\

As mentioned in the Introduction, the notion of fragility has been used as a classification scheme for glass-forming systems and has been empirically correlated 
to a variety of other properties associated with glass formation. One may then hope that if the temperature dependence of the relaxation and viscous slowdown of a 
given glass-former is characterized by a unique fragility, whatever its density, one would be able to derive more meaningful correlations with other properties.  
%\textcolor{blue}{One may also wonder if the density scaling of the relaxation time and the viscosity extends to the whole relaxation functions.
Attempts in this direction [\onlinecite{niss07}] have used the observation that the stretching exponent and more generally the shape and wavevector-dependence of the relaxation 
function appear to a good approximation to be constant along an isochronic (i.e., equal relaxation time) line [\onlinecite{Frick2003}]. However, the correlation between the 
stretching exponent and the isochoric fragility for a variety of liquids does not seem significantly improved compared to the isobaric case [\onlinecite{bohmer_stretching}] 
and does not go beyond a general trend that the stretching exponent roughly increases as fragility decreases. More favorable correlations with an isochoric fragility 
than with the standard isobaric one have been found in the case of the intensity of the Boson peak and of the nonergodicity factor [\onlinecite{niss06}]. Clearly a more 
systematic investigation would be required to revisit the empirical correlations proposed in the literature in light of the isochoric fragility.
\\

\begin{figure}[t]
\includegraphics[width=0.99\columnwidth]{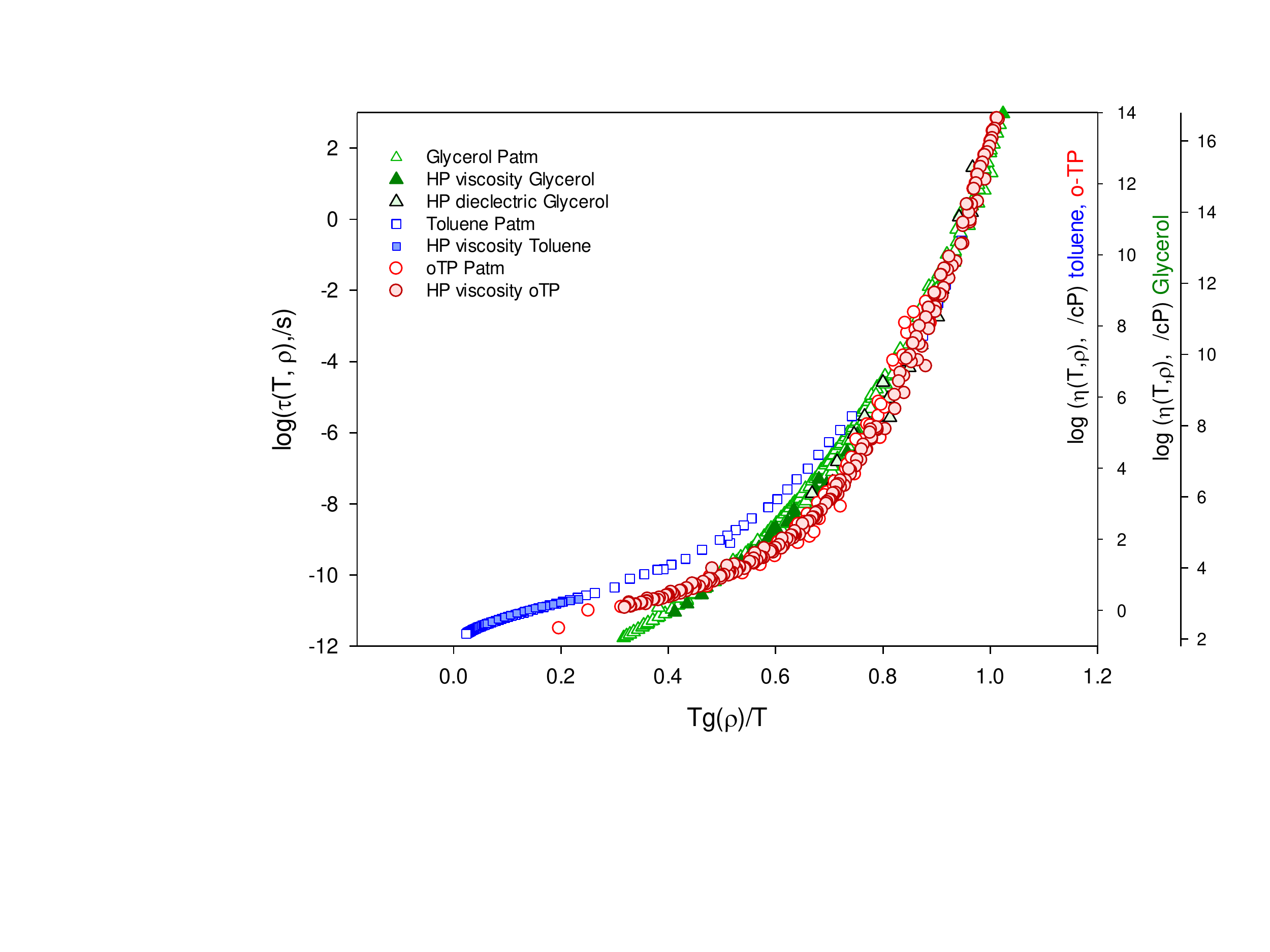}
\caption{Isochoric Angell plot for the $\alpha$-relaxation time and the viscosity of several glass-forming molecular liquids, o-terphenyl, glycerol and toluene: 
$\log_{10}(\tau(\rho,T)/s)$ or $\log_{10}(\eta(\rho,T)/cP)$ versus $T_g(\rho)/T\equiv X_g/X$ (see text). Same data as in Fig. \ref{Fig_density-scaling}. Note that 
the difference in fragility (steepness index at $T_g$) seen in Fig. \ref{Fig_Tg-scaling} between glycerol and o-terphenyl is now much reduced. The liquids appear 
to have a significantly narrower range of isochoric fragilities with  $m_\rho\sim 38-45$ (and for all liquids studied so far, $m_\rho \le m_p(P_{\rm atm})$). }
\label{Fig_isochoric_Angell}
\end{figure}

Going back to the strong-fragile representation put forward by Austen Angell via its celebrated scaled Arrhenius plot, the above property of the isochoric fragility allows us 
to build an an ``isochoric Angell plot'' in which $\log_{10}(\tau(\rho,T)/\tau_{\infty})$ is displayed as a function of $X_g/X$ with $X=e(\rho)/T$, or equivalently $T_g(\rho)/T$ 
(since, trivially, $X_g/X=T_g(\rho)/T$). We illustrate such a plot in Fig. \ref{Fig_isochoric_Angell}, where we display data collected for all available $(P,T)$ state points 
for several glass-forming molecular liquids (from the boiling point down to $T_g$).% ionic liquids, and polymers.

As can be seen from the figure, when compared to the same liquids represented in the conventional Angell plot, there is a reduction of the breadth of fragilities when passing 
from the isobaric to the isochoric ones. This is due to the fact that fragile organic liquids such as o-terphenyl are more sensitive to density than for instance the 
hydrogen-bonded liquid glycerol. This extra effect of the density on the super-Arrhenius slowdown is now scaled out when considering isochoric plots that more directly 
characterize the intrinsic role of temperature.
\\

\section{Scaling out the high-temperature behavior to compare different liquids}
\label{sec_arrhenius}

We have discussed the way to more crisply characterize the fragility of a given glass-former by scaling out the density and thereby defining a unique (isochoric) 
fragility. Another possible line of thought to improve the significance of the strong-fragile classification is to envisage somehow correcting the fragility for the contribution 
of the high-temperature, Arrhenius-like, behavior. This idea again goes back to Daniel Kivelson's work, whose take was that the observed super-Arrhenius temperature 
dependence of the relaxation time and transport coefficients of supercooled liquids at low temperature near $T_g$ results from the combination of a collective phenomenon 
by which relaxation becomes more and more cooperative as temperature decreases and a more local or molecular effect that sets the bare activation energy scale of a 
liquid [\onlinecite{kivelson95,kivelson96}]. This view is more controversial but is worth exploring. It rests on the already mentioned observation that the temperature evolution 
of the relaxation time (or viscosity) of a glass-forming liquid can be described as being Arrhenius-like at high temperature (see also above),
\begin{equation}
\label{eq_tau_arrhenius}
\tau(T)=\tau_{\infty}\, {\rm exp}\left (\frac{E_{\infty}}{T}\right )\, ,
\end{equation}
as usually found to a good approximation [\onlinecite{kivelson96,rossler}], and super-Arrhenius at low temperature, 
\begin{equation}
\label{eq_tau_superarrhenius}
\tau(T)=\tau_{\infty}\, {\rm exp}\left (\frac{E(T)}{T}\right )
\end{equation}
with $E(T)$ growing as $T$ decreases. (Note that $\tau_\infty$ is here taken as a material- and probe-dependent adjustable parameter.) One would then like to 
disentangle the genuinely collective nature of the super-Arrhenius dependence from the local activation 
energy scale that is directly associated with the nature of the bonding between nearby molecules. This led for instance to the already mentioned proposal of sorting out 
liquids not simply along  a strong-fragile axis but according to two axes [\onlinecite{ferrer_extension}]: a strong/weak axis based on the value of $E_\infty/T$ at a 
common reference (high) temperature, e.g., melting, and a fragile/nonfragile axis associated with the degree of super-Arrhenius behavior.
\\

Building on this idea, one can also try to scale out the effect of the bare activation energy $E_\infty$ and see how this affects (or not) the relative fragility of 
glass-forming liquids. In the spirit of the Angell plot, one proposal would be to study the relaxation time at temperature $T$ [see Eq. (\ref{eq_tau_superarrhenius})] divided by the extrapolated Arrhenius-like formula [see Eq. (\ref{eq_tau_arrhenius})] and to consider this ratio $\tau(T)/\tau_{\rm arrh}(T)$ as a function of a rescaled inverse temperature. 
To then scale $T$ one can choose a temperature at which $\log_{10}(\tau(T)/\tau_{\rm arrh}(T))$ is equal to some fixed value. (Of course, if the value is chosen sufficiently large 
to include a significant super-Arrhenius course, strong glass-formers with an almost Arrhenius behavior cannot be represented because 
$\log_{10}(\tau(T)/\tau_{\rm arrh}(T))\approx 0$: we therefore restrict our study to fragile and intermediate systems.) Let $K_\#$ denote the chosen value of 
$\log_{10}(\tau(T)/\tau_{\rm arrh}(T))$ and $T_\#$ denote the temperature at which this value is reached. Then, we introduce a steepness index to characterize fragility as
\begin{equation}
m_\#=\frac{\partial \log_{10}(\tau(T)/\tau_{\rm arrh}(T))}{\partial(T_\#/T)}\Big \vert_{T=T_\#}.
\end{equation}
One easily checks that this definition preserves the property that the isochoric fragility is independent of density. This can be done by again making use of the scaling form 
in Eq. (\ref{eq_rho_scaling}) and of the fact that $E_\infty(\rho)=A e(\rho)$ with $A$ a constant that depends on the precise determination of $e(\rho)$. (If 
$e(\rho)$ is directly obtained as the high-temperature Arrhenius effective activation energy, then $A=1$, whereas $A\neq 1$ if one for instance simply represents $e(\rho)$ 
by a power law $\rho^x$.) Then, one finds
\begin{equation}
m_\#(\rho)=X_\#[\mathcal F'(X_\#)-A],
\end{equation}
where $X_\#=e(\rho)/T_\#(\rho)$ satisfies $\mathcal F(X_\#)-AX_\#=K_\#$ and is thus independent of $\rho$. As anticipated, the new steepness index is independent of 
density when defined under isochoric conditions, despite the fact that the bare activation energy $E_\infty(\rho)$ varies with density.

\begin{figure}[t]
\includegraphics[width=0.99\columnwidth]{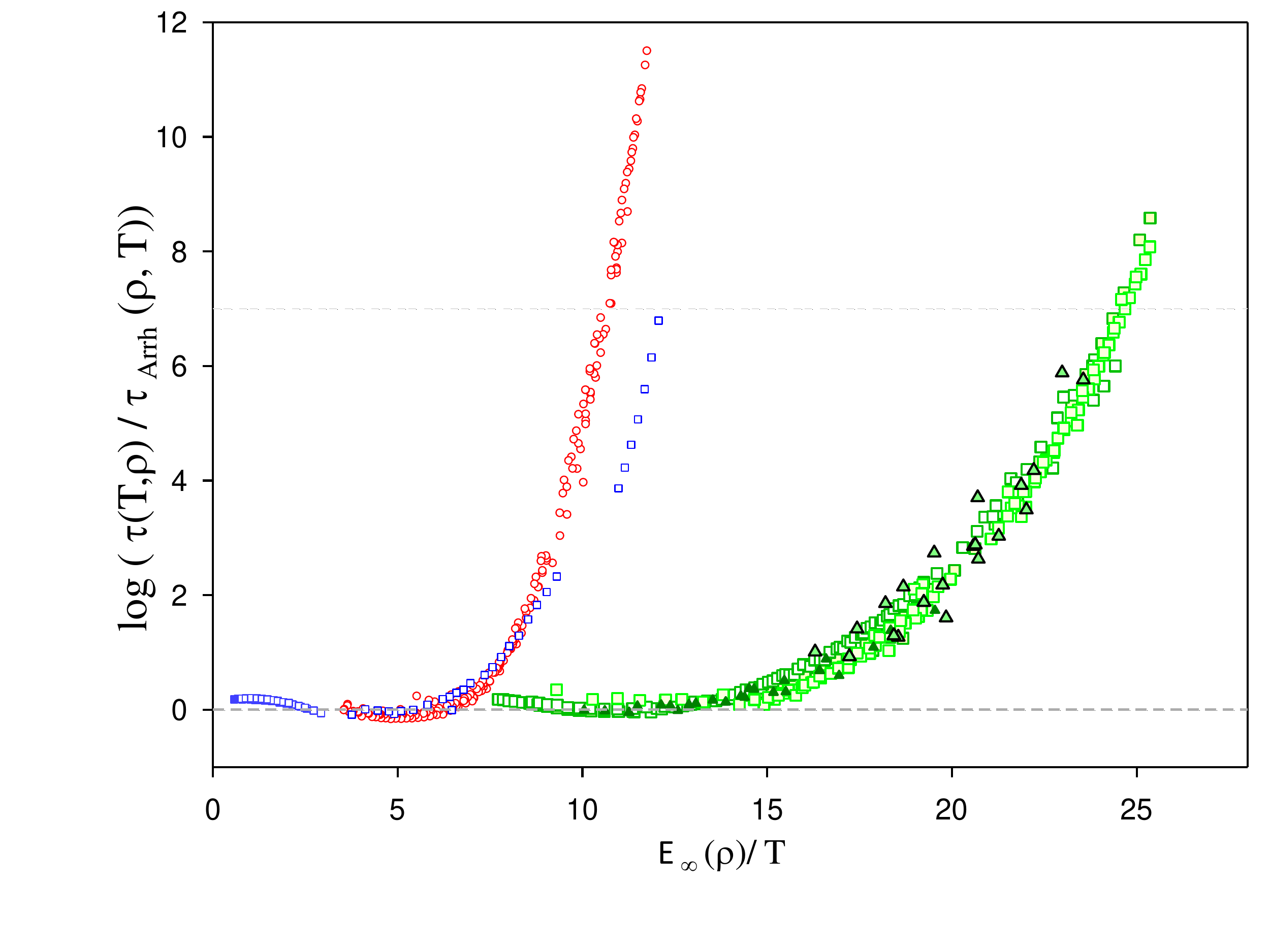}
\includegraphics[width=0.99\columnwidth]{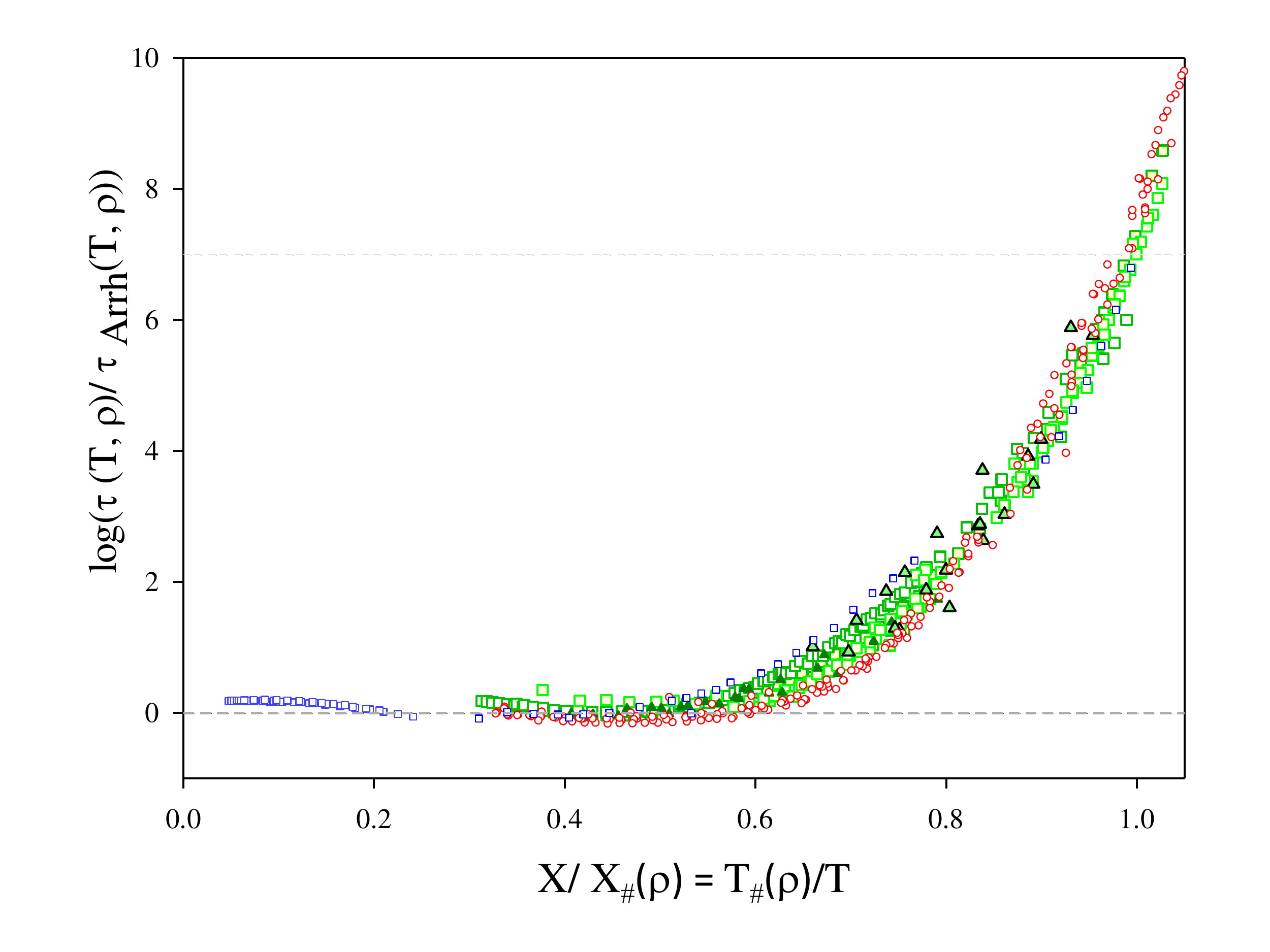}
\caption{Scaled plot of the super-Arrhenius behavior accounting for the effect of the bare activation energy. Upper:  $\log_{10}(\tau(\rho,T)/\tau_{\rm arrh}(\rho, T))$ 
versus the scaling variable $X=E_\infty(\rho)/T\propto e(\rho)/T$ (see text). Lower:  Angell-like plot versus $X/X_\#=T_\#(\rho)/T$.  $X_\#$ (or $T_\#$) is 
chosen such that $\log_{10}(\tau(\rho,T)/\tau_{\rm arrh}(\rho, T))$ reaches the common value of 7 that is indicated by the dotted line in the upper panel. (Data on toluene, 
o-terphenyl, and glycerol are taken from the same references as for Fig. \ref{Fig_density-scaling}  and \ref{Fig_isochoric_Angell}.)}
\label{Fig_arrhenius_rescaled}
\end{figure}

Although it is not always possible to collect enough experimental data to reconstruct the isochoric plots and the determination of an Arrhenius effective activation energy is 
a somehow ambiguous (and not necessarily justified at very low densities or close to the boiling point) procedure, we illustrate the outcome with three different glass-forming 
molecular liquids, already shown in the previous figures. In the upper panel of Fig. \ref{Fig_arrhenius_rescaled} we display 
$\log_{10}(\tau(\rho,T)/\tau_{\rm arrh}(\rho, T))$ versus the scaling variable $X=E_\infty(\rho)/T$ and in the lower panel we show the same data 
versus $X/X_\#=T_\#(\rho)/T$ where $X_\#$ (and  then $T_\#$) is chosen such that $\log_{10}(\tau(\rho,T)/\tau_{\rm arrh}(\rho, T))$ reaches the fixed value of $7$ to include 
all experimental data at our disposal without extrapolations. (One might expect that more data in the high density/high relaxation time regime will be available in the future.) 
This new representation amounts to a modified isochoric Angell plot in which both the density and the high-temperature effective activation energy are scaled out.

We can see that the behavior of the three liquids that have already been considered in Fig. \ref{Fig_isochoric_Angell} appears even closer in the new plot, showing a 
further reduction of the difference in fragility. This seems to indicate that when the effect of the density and that of the liquid-specific bonding, which is reflected in the 
high-temperature effective activation energy, are scaled out to unveil the genuine super-Arrhenius character, liquids that seemed quite different in terms of the conventional 
fragility criterion (see Fig. \ref{Fig_Tg-scaling} Right) are actually very similar. (A note of caution must be given: the amplitude of the variation is more limited in 
Fig. \ref{Fig_arrhenius_rescaled} than in Fig. \ref{Fig_Tg-scaling} or \ref{Fig_isochoric_Angell}, so that differences could possibly build up if one were able to cover a wider 
range for $\log_{10}(\tau(T)/\tau_{\rm arrh}(T))$.) It would be interesting to obtain data on a broad spectrum of glass-forming liquids to check if the trends that we have 
observed in the present study are general.
\\

Note that an alternative procedure for trying to remove part of the effect of the bare or molecular contribution to the effective activation energy is to scale 
the temperature by a ``high'' reference temperature in place of $T_g$ or of the above $T_\#$. This reference temperature can be taken as the crossover 
or onset temperature $T_*$ at which deviation from Arrhenius behavior starts being detectable. The choice of such a high-$T$ crossover point has the 
merit of allowing a direct comparison of simulation and laboratory data [\onlinecite{fragilityLJ,coslovich}] and also easily leads  to a subtraction of the high-$T$ effect 
in order to define an intrinsic fragility that do not depend on an arbitrary time scale: this is what is done in the frustration-limited domain approach of glass 
formation [\onlinecite{kivelson95,kivelson96}] and can be also achieved by studying the function $(E(T)-E_{\infty})/E_{\infty}$ [\onlinecite{alba02,tarjus04}]. 
However, finding a robust operational way for defining the crossover is far from trivial.
\\

One should finally stress that the very idea of scaling out a bare activation energy associated with some fairly universal Arrhenius-like regime at high temperature 
is not unanimously accepted. The dynamics of simple liquids, such as Argon, at high temperature has no reason to behave according to an activated 
Arrhenius picture [\onlinecite{chandler}]. The situation may however be different for glass-forming liquids whose relaxation and transport properties seem to be described 
by significant effective activation energies even above melting [\onlinecite{kivelson96,rossler}]. In any case, one may avoid any reference to a bare activation energy or 
include it in the general theoretical description of the relaxation slowdown. This is what is done for instance in the phenomenological extension of the mean-field theory 
of glass formation known as the random first-order transition (RFOT) theory [\onlinecite{KWT89,lubchenko-wolynes}]. One can also decide to define fragility with an 
unreachable reference temperature that is below $T_g$. This is in effect what is done when using the VTF formula to describe the temperature dependence of the 
relaxation time (or the viscosity),
\begin{equation}
\tau(T)=\tau_0 \exp \left (D \frac{T_0}{T-T_0} \right )
\end{equation}
with $T_0<T_g$ and $D$ an intrinsic measure of fragility (fragility decreases as $D$ increases). The VTF formula was for instance used for illustration by Austen Angell 
when introducing the strong/fragile classification [\onlinecite{angell91}]. In this case, the steepness 
index at $T_g$, $m_g$, is approximately related to the coefficient $D$ through $m_g\approx 17(1+37/D)$, a fragile system having a small $D$ and a large $m_g$ 
and conversely a strong system having a large $D$ and a small $m_g$. The RFOT approach predicts that $D\propto 1/\Delta c_p$ where $\Delta c_p$ is the heat 
capacity jump per mole at $T_g$ [\onlinecite{xia-wolynes}]. This is in line with Angell's first arguments about fragility [\onlinecite{angell85,angell85_bis}].
Note that when $T\gg T_0$ one formally recovers from the VTF formula an Arrhenius behavior (although it is very generally found that the VTF fit no longer works at 
high temperature: see, e.g., [\onlinecite{kivelson96,stickel95,rossler}]) with $E_\infty=D T_0/\ln(10)$. In this case the steepness index $m_\#$ is also uniquely determined by 
$D$ as $m_g$, with the same trend. This is due to the direct link between $E_\infty/T_g$ and $E_\infty/T_\#$ and the fragility index $1/D$.
\\

\section{Conclusion: Is fragility connected to ``cooperativity'' and collective behavior?}

As we have already alluded to, the above developments about pruning the notion of fragility from extraneous effects have a strong motivation in the theoretical view 
that since fragility characterizes the degree of stronger-than-Arrhenius temperature dependence of the relaxation and viscous slowdown it can be taken as a signature 
of a growing collective or cooperative behavior as temperature decreases. (Note that this is also why we focus on molecular liquids and do not consider the strong 
network-forming systems.) This would further justify the existence of correlations among quantities similarly reflecting this collective behavior. Needless to say that this 
view is not unanimously accepted. If anyhow one pursues in this direction, one should detail a bit more what goes under the idea of a cooperative and/or 
collective character of glass formation.

``Cooperativity'' in the context of thermally activated dynamics means that degrees of freedom must conspire to make the relaxation possible (or faster than 
by other means). In consequence, the effective barrier to relaxation is determined by the minimum number of degrees of freedom that are cooperatively 
involved. This idea, which has been made popular by Adam and Gibbs with their notion of  ``cooperatively rearranging regions'' [\onlinecite{adam-gibbs65}], is at the 
core of several theoretical approaches of the glass transition [\onlinecite{lubchenko-wolynes,BB_review,tarjus_review}]. It is of course a form of collective behavior, 
which in this case is rooted in the statics of the glass-formers, as in the Adam-Gibbs approach where the size of the cooperatively rearranging regions goes 
inversely with the configurational entropy.

How does one relate the super-Arrhenius dependence of the relaxation time to a growing static length? A heuristic derivation rests on the following assumptions:

(i) If a system has a finite correlation length it can be divided into independent subsystems of size larger than (but of the order of) this correlation length. In the 
absence of any obvious form of order in glass-forming liquids, the relevant (static) length should describe how far a condition at the boundary can influence 
the interior of the subsystem. This corresponds to a ``point-to-set'' correlation length $\xi_{\rm pts}$ [\onlinecite{BB_PTS,cavagna_PTS,montanari06,sho_PTS}].

(ii) A finite-size subsystem has a finite relaxation time whose magnitude can be related to its size. Assuming that the relaxation is thermally activated the most 
expensive activation barrier  then goes as the volume of the subsystem, as hypothesized by Adam and Gibbs.

Then, the relaxation time of the full system is bounded from above as
\begin{equation}
\label{eq_time_PTS}
\log[\tau(T)/\tau_{\infty}]\leq \frac{A}{T}N_{CRR}(T)
\end{equation}
where $N_{\rm CRR}\propto \xi_{\rm pts}^d$. The bound can be made rigorous under some conditions [\onlinecite{montanari06}]. Note that it is more likely that activation 
proceeds via lower barriers that scale with the length $\xi_{\rm pts}$ with an exponent $\psi<d$ (even faster relaxation mechanisms are also possible: see, e.g.,  
[\onlinecite{wyart-cates}]).

Assuming that the number of correlated molecules (or beads [\onlinecite{lubchenko-wolynes}]) goes to $1$ when the relaxation is no longer cooperative and the relaxation 
becomes Arrhenius-like, one can rewrite the above equation as
\begin{equation}
\label{eq_NCRR}
N_{\rm CRR}(T)\geq \frac{\log_{10} \left (\tau(T)/\tau_\infty \right )}{\log_{10} \left (\tau_{\rm arrh}(T)/\tau_\infty \right )},
\end{equation}
which make the connection with fragility and the discussion in Sec. \ref{sec_arrhenius}.
\\

Another candidate of collective behavior comes with the phenomenon of dynamic heterogeneity [\onlinecite{ediger_DH,DH_book}]. The dynamics of the molecules become spatially 
correlated over longer distances as the relaxation becomes slower, and, at a given temperature, the extent of the correlation is maximum for a time of the order 
of the $\alpha$-relaxation time. This property is more crisply captured by studying $4$-point space-time correlation functions, 
$<\delta c({\bf 0}; 0,t=\tau(T))\delta c({\bf r};0,t=\tau(T))>$, where $c({\bf r};0,t)$ characterizes the dynamics between times $0$ and $t$ around point ${\bf r}$ and 
$\delta c$ is its fluctuation [\onlinecite{dynhet,dynhet_bis}]. 
The maximum number of dynamically correlated molecules can then be lower bounded as follows [\onlinecite{dalle-ferrier07,dynhet}]
\begin{equation}
\begin{aligned}
\label{eq_Ndyn}
N_{\rm dyn}(T) &= \rho \int d^dr <\delta c({\bf 0}; 0,t=\tau(T))\delta c({\bf r};0,t=\tau(T))>
\\& \geq \frac{1}{T^2 c_p}\Phi'(1)^2\left [\frac{\partial \ln (\tau(T)/\tau_\infty)}{\partial (1/ T)}\right ]^2
\end{aligned}
\end{equation}
where $c_p$ is the heat capacity at constant pressure and $\Phi(t/\tau(T))\equiv <(1/V)\int d^drc({\bf r};0,t)>$ is the normalized relaxation function  and we have 
neglected for simplicity the small variation of its stretching with temperature.

Being less rigorous, one may still expect to obtain a bound in the following form [\onlinecite{dalle-ferrier07}]
\begin{equation}
N_{\rm dyn}(T)\gtrsim \frac{\beta^2}{T^2 \Delta c_p}\left [\frac{\partial \ln (\tau(T)/\tau_\infty)}{\partial (1/ T)}\right ]^2
\end{equation}
where $\beta$ is the stretching exponent which is taken as constant and $\Delta c_p$ is the heat-capacity jump at $T_g$.

The above expressions are given at constant pressure but they are easily generalized at constant density by making use of the density scaling of 
the dynamics. From the data in [\onlinecite{tarjus04}] we find that the lower bound for $N_{\rm CRR}(T)$ grows as $T$ decreases, but 
by only a small amount: it goes from $1$ at high temperature (by construction) to about $3$ at $T_g$ ($3.4$ for o-terphenyl and $2.7$ for glycerol). This small 
growth, less than reported in actual calculations [\onlinecite{BB_PTS,cavagna_PTS,sho_PTS,berthier_ceiling}] is likely due to the fact that 
we are only measuring a lower bound but it is nonetheless in line with the outcome of fits to the Adam-Gibbs formula and of experimental measurements [\onlinecite{ladieu}]. 
In contrast, as shown in [\onlinecite{dalle-ferrier07}], the variation of the lower bound for  $N_{\rm dyn}(T)$ is clearly larger but does not seem to strongly correlate 
with fragility. So, while increasing fragility seems to correspond to (weakly) increasing cooperativity, the spatial extent of the dynamical heterogeneity as temperature 
decreases is only remotely related with fragility. In fact, as can be seen from Eqs. (\ref{eq_NCRR}) and (\ref{eq_Ndyn}), a purely Arrhenius behavior is associated 
(by construction, here) to the absence of cooperative effect whereas the number of dynamically correlated molecules increases as $1/T^2$ when temperature 
decreases. This of course must be taken with a grain of salt and further investigated.

\begin{acknowledgements}
We thank E. R\"ossler for providing us with his experimental data on many molecular glass-forming liquids up to the boiling temperature.
\end{acknowledgements}

\end{document}